\documentclass[3p,times,twocolumn,fleqn]{elsarticle}
 \biboptions{comma,sort&compress}
 
\usepackage{graphicx}
\usepackage{here}
\usepackage{ecrc}

\usepackage{slashed}
\usepackage{bm}
\usepackage{float}

\volume{00}

\firstpage{1}

\journalname{Nuclear and Particle Physics Proceedings}

\runauth{Marco Frasca}

\jid{nppp}

\jnltitlelogo{Nuclear and Particle Physics Proceedings}

\usepackage{amssymb}

\usepackage[figuresright]{rotating}

\begin{document}

\begin{frontmatter}

\title{ Nambu-Jona-Lasinio model correlation functions from QCD }
 \cortext[cor0]{Talk given at 24th International Conference in Quantum Chromodynamics (QCD 21),  5 - 9 July 2021, Montpellier - FR}
 \author[label1]{Marco Frasca\fnref{fn1}}
 \fntext[fn1]{Speaker, Corresponding author.}
\ead{marcofrasca@mclink.it}
\address[label1]{Via Erasmo Gattamelata, 3, 
00176 Rome (Italy)}
\author[label2]{Anish Ghoshal}
\address[label2]{INFN, Rome, Italy and Warsaw University, Poland}
\author[label3]{Stefan Groote}
\address[label3]{University of Tartu, Estonia}


\pagestyle{myheadings}
\markright{ }
\begin{abstract}
We treat quantum chromodynamics (QCD) using a set of Dyson-Schwinger equations derived, in differential form, with the Bender-Milton-Savage technique. In this way, we are able to derive the low energy limit that assumes the form of a non-local Nambu-Jona-Lasinio model with all the parameters properly fixed by the QCD Lagrangian and the determination of the mass gap of the gluon sector.
\end{abstract}

\end{frontmatter}
\section{Introduction}

Adding quarks to the Yang-Mills Lagrangian makes the theory not exactly treatable.
Notwithstanding such a difficulty, full QCD can be handled with Dyson-Schwinger equations even if some approximations are needed to get the low-energy limit.
Such approximations entail both the strong coupling limit and 't Hooft limit $N\rightarrow\infty,\ \ Ng^2=constant\gg 1$.
In this way, we will recover, in the low-energy limit, the equations for the correlation functions of a non-local NJL model.

Yang-Mills set of Dyson-Schwinger equations can be solved through a class of exact solutions of the 1-point function, similarly to the $\phi^4$ theory \cite{Frasca:2015yva}. The spectrum can be obtained as well and appears in agreement with lattice data \cite{Frasca:2017slg}. Adding quarks to the theory makes it unsolvable but, also in this case, our approach could be proven meaningful to derive the proper low energy limit of QCD.

Our idea is to provide a method to derive the full hierarchy of Dyson-Schwinger equations also for QCD, retaining their full differential form. This is possible provided we use a technique devised by Bender, Milton and Savage \cite{Bender:1999ek}.

Some approximations are needed to get the low-energy limit.
Such approximations entail both the strong coupling limit and 't Hooft limit $N\rightarrow\infty$ and $Ng^2=constant\gg 1$ \cite{tHooft:1973alw,tHooft:1974pnl}.
In this way, we will recover, in the low-energy limit, the equations for the correlation functions of a non-local Nambu-Jona-Lasinio model.

This is an interesting result in view of the fact that it permits to recover from an error present in Ref.~\cite{Frasca:2019ysi} granting anyway the conclusions. This has been recently applied to the $g-2$ problem \cite{Frasca:2021yuu} to evaluate the hadronic vacuum polarization for the $\pi\pi$ contribution.

\section{Bender-Milton-Savage technique}


This technique can be better explained referring to a scalar field. We consider the following partition function
\begin{equation}
    Z[j]=\int[D\phi]e^{iS(\phi)+i\int d^4xj(x)\phi(x)}.
\end{equation}
To derive 1P-function, one has
\begin{equation}
\left\langle\frac{\delta S}{\delta\phi(x)}\right\rangle=j(x),
\end{equation}
assuming
\begin{equation}
\left\langle\ldots\right\rangle=\frac{\int[D\phi]\ldots e^{iS(\phi)+i\int d^4xj(x)\phi(x)}}{\int[D\phi]e^{iS(\phi)+i\int d^4xj(x)\phi(x)}}
\end{equation}
By setting $j=0$ one obtains the equation for th 1P-function. Next, we derive this equation again with respect to $j$ to get the equation for the 2P-function. We are taking for the nP-functions the following definition
\begin{equation}
\langle\phi(x_1)\phi(x_2)\ldots\phi(x_n)\rangle=\frac{\delta^n\ln(Z[j])}{\delta j(x_1)\delta j(x_2)\ldots\delta j(x_n)}.
\end{equation}
This implies
\begin{equation}
\frac{\delta G_k(\ldots)}{\delta j(x)}=G_{k+1}(\ldots,x).
\end{equation}

Such a procedure can be iterated to whatever order giving, in principle, all the hierarchy of the Dyson-Schwinger equations in PDE form. Going to higher orders could imply complicated computations but this approach shows itself to be very useful when some known solutions are given for 1P- and 2P-functions as in our case.

\section{1P and 2P functions for QCD}

We work in the Landau gauge as can make some computations simpler as seen in \cite{Frasca:2015yva}.

The Bender-Milton-Savage method yields for the 1P-functions
\begin{eqnarray}
      &&\partial^2G_{1\nu}^{a}(x)+gf^{abc}(
		\partial^\mu G_{2\mu\nu}^{bc}(0)+ \nonumber \\
		&&\partial^\mu G_{1\mu}^{b}(x)G_{1\nu}^{c}(x)-
		\partial_\nu G_{2\mu}^{\nu bc}(0)
		\nonumber \\
		&&-\partial_\nu G_{1\mu}^{b}(x)G_{1}^{\mu c}(x)) \nonumber \\
		&&+gf^{abc}\partial^\mu G_{2\mu\nu}^{bc}(0)+gf^{abc}\partial^\mu(G_{1\mu}^{b}(x)G_{1\nu}^{c}(x))
		\nonumber \\
		&&+g^2f^{abc}f^{cde}(G_{3\mu\nu}^{\mu bde}(0,0)
		+G_{2\mu\nu}^{bd}(0)G_{1}^{\mu e}(x)
		\nonumber \\
	    &&+G_{2\nu\rho}^{eb}(0)G_{1}^{\rho d}(x)
	    +G_{2\mu\nu}^{de}(0)G_{1}^{\mu b}(x)+ \nonumber \\
	    &&G_{1}^{\mu b}(x)G_{1\mu}^{d}(x)G_{1\nu}^{e}(x)) \nonumber \\
		&&=g\sum_{q,i}\gamma_\nu T^aS_{q}^{ii}(0)+g\sum_{q,i}{\bar q}_1^i(x)\gamma_\nu T^a q_1^i(x),
\end{eqnarray}
and for the quarks
\begin{equation}
	(i\slashed\partial-m_q)q_{1}^{i}(x)+g{\bm T}\cdot\slashed{\bm G}_1(x) q_{1}^{i}(x) 
	+g{\bm T}\cdot\slashed{\bm W}^{i}_q(x,x)= 0.
\end{equation}

Here and in the following Greek indexes ($\mu,\nu,\ldots$) are for the space-time and Latin index ($a, b,\ldots$) for the gauge group. A usual for the Dyson-Schwinger set, the equations for the lower order nP-functions depend on the higher order correlation functions. We will see how to treat this aspect in the following.

At this stage, we apply the re-mapping idea to such equations as done in \cite{Frasca:2015yva}. So, we assume 
\begin{equation}
G_{1\nu}^a(x)\rightarrow\eta_\nu^a\phi(x)
\end{equation}
being $\phi(x)$ a scalar field. Let us introduce the $\eta$-symbols as follows
\begin{eqnarray}
\eta_\mu^a\eta^{a\mu} &=& N^2-1. \nonumber \\ 
\eta_\mu^a\eta^{b\mu} &=& \delta_{ab}, \nonumber \\
\eta_\mu^a\eta_\nu^a &=& \left(g_{\mu\nu}-\delta_{\mu\nu}\right)/2.
\end{eqnarray}
All this permits to get the reduced equations
\begin{eqnarray}
&&\partial^2\phi(x)+2Ng^2\Delta(0)\phi(x)+Ng^2\phi^3(x) 
\nonumber \\
&&=\frac{1}{N^2-1}\left[g\sum_{q,i}\eta^{a\nu}\gamma_\nu T^aS_{q}^{ii}(0)\right. \nonumber \\
&&\left.+g\sum_{q,i}{\bar q}_1^i(x)\eta^{a\nu}\gamma_\nu T^a q_1^i(x)\right]
\nonumber \\
&&(i\slashed\partial-m_q^i)q_{1}^{i}(x)+g{\bm T}\cdot\slashed\eta\phi(x) q_{1}^{i}(x) = 0.
\end{eqnarray}

We do the same for the 2P-functions. In the Landau gauge, the gluon 2P-function takes the form 
\begin{equation}
    G_{2\mu\nu}^{ab}(x-y)=\left(\eta_{\mu\nu}-\frac{\partial_\mu\partial_\nu}{\partial^2}\right)\Delta_\phi(x-y)
\end{equation}
being $\eta_{\mu\nu}$ the Minkowski metric, and $\Delta_\phi(x-y)$ is the propagator of the $\phi$ given the map between the scalar and the Yang-Mills fields. Finally, we can write
\begin{eqnarray}
&&\partial^2\Delta_\phi(x-y)+2Ng^2\Delta_\phi(0)\Delta_\phi(x-y)+3Ng^2\phi^2(x)\Delta_\phi(x-y) \nonumber \\
&&=g\sum_{q,i}{\bar Q}^{ia}_\nu(x-y)\gamma^\nu T^a q_{1}^{i}(x)
\nonumber \\
&&+g\sum_{q,i}{\bar q}_1^{i}(x)\gamma^\nu T^a Q^{ia}_\nu(x-y) + \delta^4(x-y)\nonumber \\
&&\partial^2 P^{ad}_2(x-y)=\delta_{ad}\delta^4(x-y) \nonumber \\
&&(i\slashed\partial-m_q^i)S^{ij}_q(x-y) \nonumber \\
&&+g{\bm T}\cdot\slashed\eta\phi(x) S^{ij}_q(x-y)=\delta_{ij}\delta^4(x-y) \nonumber \\  
&&\partial^2W_{q\nu}^{ai}(x-y)+2Ng^2\Delta_\phi(0)W_{q\nu}^{ai}(x-y)+3Ng^2\phi^2(x)W_{q\nu}^{ai} \nonumber \\
&&=g\sum_{j}{\bar q}_1^{j}(x)\gamma_\nu T^a S^{ji}_q(x-y)\nonumber \\
&&(i\slashed\partial-m_q^i)Q^{ia}_\mu(x-y)+g{\bm T}\cdot\slashed\eta\phi(x) Q^{ia}_\mu(x-y) \nonumber \\
&&+gT^a\gamma_\mu\Delta_\phi(x-y) q_{1}^{i}(x)=0.
\end{eqnarray}

\section{'t Hooft limit}

't Hooft limit means to solve the theory assuming \cite{tHooft:1973alw, tHooft:1974pnl}
\begin{equation}
  N\rightarrow\infty,\qquad Ng^2=constant, \qquad Ng^2\gg 1.
\end{equation}
In our case, the gauge group is SU(N) being $N$ is the number of colors. To evaluate our equations in such a limit, we need a perturbation series for a very large coupling. We proposed such a technique in Ref.\cite{Frasca:2013tma}. We do a rescaling, $x\rightarrow\sqrt{Ng^2}x$, and write the equation for the gluon field as follows
\begin{eqnarray}
      \partial^2\phi(x')+2\Delta_\phi(0)\phi(x')+3\phi^3(x')&=& \\
\frac{1}{\sqrt{Ng^2}\sqrt{N}(N^2-1)}\left[\sum_{q,i}\eta\cdot\gamma\cdot TS_{q}^{ii}(0)+\right.&& \nonumber \\
\left.\sum_{q,i}{\bar q}_1^i(x')\eta\cdot\gamma\cdot T q_1^i(x')\right].&&\nonumber
\end{eqnarray}
In the 't Hooft limit, we get at the leading order for the 1P-functions
\begin{eqnarray}
	\partial^2\phi_0(x)+2Ng^2\Delta_\phi(0)\phi_0(x)+3Ng^2\phi_0^3(x)=0,& \nonumber \\
		(i\slashed\partial-m_q^i){\hat q}_{1}^{i}(x)+g{\bm T}\cdot\slashed{\eta}\phi(x) q_{1}^{i}(x)=0.&
\end{eqnarray}
At the leading order the only effect is seen on masses. For the quark field, this will be clearer below. We can solve the equation for the gluon field taking
\begin{eqnarray}
\phi_0(x)=\sqrt{\frac{2\mu^4}{m^2+\sqrt{m^4+2Ng^2\mu^4}}}\times
\nonumber \\
{\rm sn}\left(p\cdot x+\chi,\kappa\right),
\end{eqnarray}
being sn a Jacobi elliptical function, $\mu$ and $\chi$ arbitrary integration constants and $m^2=2Ng^2\Delta_\phi(0)$. We have
\begin{equation}
\kappa=\frac{-m^2+\sqrt{m^4+2Ng^2\mu^4}}{-m^2-\sqrt{m^4+2Ng^2\mu^4}}.
\end{equation}
This is true provided that the following dispersion relation holds
\begin{equation}
    p^2=m^2+\frac{Ng^2\mu^4}{m^2+\sqrt{m^4+2Ng^2\mu^4}}.
\end{equation}
For the equations of the 2P-functions one has
\begin{eqnarray}
\partial^2\Delta_\phi(x,y)+2Ng^2\Delta_\phi(0)\Delta(x-y)+3Ng^2\phi_0^2(x)\Delta_\phi(x-y) \nonumber \\
=g\sum_{q,i}{\bar Q}^{ia}_\nu(x,y)\gamma^\nu T^a {\hat q}_{1}^{i}(x)
\nonumber \\
+g\sum_{q,i}{\bar{\hat q}}_1^{i}(x)\gamma^\nu T^a Q^{ia}_\nu(x,y)
+ \delta^4(x-y) \nonumber \\
\partial^2 P^{ad}_2(x-y)=\delta_{ad}\delta^4(x-y) \nonumber \\
(i\slashed\partial-m_q^i){\hat S}^{ij}_q(x-y)+g{\bm T}\cdot\slashed\eta\phi(x) S^{ij}_q(x-y)=\delta_{ij}\delta^4(x-y) \nonumber \\  
\partial^2W_{q\nu}^{ai}(x,y)+2Ng^2\Delta_\phi(0)W_{q\nu}^{ai}(x,y)+3Ng^2\phi_0^2(x)W_{q\nu}^{ai}(x,y)\nonumber \\
=g\sum_{j}{\bar {\hat q}}_1^{j}(x)\gamma_\nu T^a {\hat S}^{ji}(x-y) \nonumber \\
(i\slashed\partial-{\hat M}_q^i){\hat Q}^{ia}_\mu(x,y)+gT^a\gamma_\mu\Delta_\phi(x-y) {\hat q}_{1}^{i}(x)=0.
\end{eqnarray}
To solve these equations, let us consider
\begin{eqnarray}
\partial^2\Delta_0(x-y)+[m^2+3Ng^2\phi_0^2(x)]\Delta_0(x-y)&=&
\nonumber \\
\delta^4(x-y).&&
\end{eqnarray}
In momenta space, the solution of this equation is given by \cite{Frasca:2015yva,Frasca:2013tma}
\begin{eqnarray}
   \Delta_0(p)=M{\hat Z}(\mu,m,Ng^2)\frac{2\pi^3}{K^3(\kappa)}\times \nonumber \\
	\sum_{n=0}^\infty(-1)^n\frac{e^{-(n+\frac{1}{2})\pi\frac{K'(\kappa)}{K(\kappa)}}}
	{1-e^{-(2n+1)\frac{K'(\kappa)}{K(\kappa)}\pi}}\times \nonumber \\
	(2n+1)^2\frac{1}{p^2-m_n^2+i\epsilon}
\end{eqnarray}
being
\begin{equation}
M=\sqrt{m^2+\frac{Ng^2\mu^4}{m^2+\sqrt{m^4+2Ng^2\mu^4}}},
\end{equation}
and ${\hat Z}(\mu,m,Ng^2)$ a given constant. The spectrum is given by $m_n$ and a proper gap equation \cite{Frasca:2017slg}.

\section{QCD in the low energy limit}

Using the technique devised in \cite{Frasca:2013tma}, the next-to-leading order term is given by
\begin{eqnarray}
\phi_1(x)=g\frac{1}{N^2-1}\int d^4x'\Delta_0(x-x')
\left[\sum_{q,i}\eta\cdot\gamma\cdot TS_{q}^{ii}(0)\right. & \nonumber \\
\left.+\sum_{q,i}{\bar q}_1^i(x')\eta\cdot\gamma\cdot T q_1^i(x')\right].
\end{eqnarray}
Given equation for the quark 1P-function
\begin{equation}
(i\slashed\partial-m_q)q_{1}^{i}(x)+g{\bm T}\cdot\slashed\eta\phi(x) q_{1}^{i}(x) = 0,
\end{equation}
one has
\begin{eqnarray}
(i\slashed\partial-m_q)q_{1}^{i}(x)+g{\bm T}\cdot\slashed\eta\phi_0(x) q_{1}^{i}(x)& \nonumber \\ 
+g^2\frac{1}{N^2-1}\int d^4x'\Delta_0(x-x')
\sum_{q,k}{\bar q}_1^k(x'){\bm T}\cdot\slashed\eta q_1^k(x')\times& \nonumber \\
{\bm T}\cdot\slashed\eta q_{1}^{i}(x)
= 0.&
\end{eqnarray}
't Hooft limit implies that the $\phi_0$ term can be neglected with respect to the second one and we get
\begin{eqnarray}
(i\slashed\partial-m_q)q_{1}^{i}(x)& \nonumber \\ 
+g^2\frac{1}{N^2-1}\int d^4x'\Delta_0(x-x')
\sum_{q,k}{\bar q}_1^k(x'){\bm T}\cdot\slashed\eta q_1^k(x')\times& \nonumber \\
{\bm T}\cdot\slashed\eta q_{1}^{i}(x)
= 0.& 
\end{eqnarray}
For the quark propagator one has instead
\begin{equation}
(i\slashed\partial-m_q)S^{ij}_q(x-y)+g{\bm T}\cdot\slashed\eta\phi(x) S^{ij}_q(x-y)=\delta_{ij}\delta^4(x-y)
\end{equation}
therefore
\begin{eqnarray}
(i\slashed\partial-m_q)S^{ij}_q(x-y)& \nonumber \\
+g{\bm T}\cdot\slashed\eta\phi_0(x) S^{ij}_q(x-y)& \nonumber \\ 
+g^2\frac{1}{N^2-1}\int d^4x'\Delta_0(x-x')
\sum_{q,k}{\bar q}_1^k(x'){\bm T}\cdot\slashed\eta q_1^k(x')\times& \nonumber \\
{\bm T}\cdot\slashed\eta S^{ij}_q(x-y)
= \delta_{ij}\delta^4(x-y).&
\end{eqnarray}
Again, by the 't Hooft limit we can neglect the $\phi_0$ term with respect to the second one and we get
\begin{eqnarray}
(i\slashed\partial-m_q)S^{ij}_q(x-y)& \nonumber \\ 
+g^2\frac{1}{N^2-1}\int d^4x'\Delta_0(x-x')
\sum_{q,k}{\bar q}_1^k(x'){\bm T}\cdot\slashed\eta q_1^k(x')\times& \nonumber \\
{\bm T}\cdot\slashed\eta S^{ij}_q(x-y)& \nonumber \\
=\delta_{ij}\delta^4(x-y).& 
\end{eqnarray}
We can recognize here the equations for the 1P- and 2P-functions of a non-local Nambu-Jona-Lasinio model. These are not generally treatable. They should be solved straightforwardly being already quantum averaged. In order to obtain a gap equation, we need to recover the Nambu-Jona-Lasinio-model from which they can be obtained
doing some kind of backtracking. Only in this way a gap equation is derived.
Indeed, such a model has the Lagrangian
\begin{eqnarray}
L_{NJL}=\sum_{i,q}\left[{\bar q}_i(x)(i\slashed\partial-m_q)q_{i}(x)+\right.
& \nonumber \\
\left.\frac{g^2}{N^2-1}{\bar q}_i(x)\int d^4x'\Delta_0(x-x')
\sum_{k,q'}{\bar q'}_k(x'){\bm T}\cdot\slashed\eta q'_k(x')
{\bm T}\cdot\slashed\eta q_{i}(x)\right].
\end{eqnarray}
From this, one can get a quark gap equation that is identical to the one given in \cite{Frasca:2019ysi} as proven in \cite{Frasca:2021yuu}.

\section{Conclusions}

We derived the set of Dyson-Schwinger equations, till to 2P-functions, for QCD with the Bender-Milton-Savage technique. We treated them in the 't Hooft limit. The low-energy limit is a nonlocal-Nambu-Jona-Lasinio model. This was shown obtaining the corresponding 1P- and 2P-equations for its correlation functions. The Lagrangian of the model is also given.

Work is ongoing to get the gap equation and to analyze its properties in view of recent g-2 Fermilab measurement \cite{Frasca:2021yuu}.

\section*{Acknowledgements\label{Ack}}
The research was supported in part by the European Regional Development Fund
under Grant No.~TK133.


\begin{thebibliography}{999}
\bibitem{Frasca:2015yva}
M.~Frasca,
Eur. Phys. J. Plus \textbf{132}, no.1, 38 (2017)
[erratum: Eur. Phys. J. Plus \textbf{132}, no.5, 242 (2017)]
doi:10.1140/epjp/i2017-11321-4
[arXiv:1509.05292 [math-ph]].

\bibitem{Frasca:2017slg}
M.~Frasca,
Nucl. Part. Phys. Proc. \textbf{294-296}, 124-128 (2018)
doi:10.1016/j.nuclphysbps.2018.02.005
[arXiv:1708.06184 [hep-ph]].

\bibitem{Bender:1999ek}
C.~M.~Bender, K.~A.~Milton and V.~Savage,
Phys. Rev. D \textbf{62}, 085001 (2000)
doi:10.1103/PhysRevD.62.085001
[arXiv:hep-th/9907045 [hep-th]].

\bibitem{tHooft:1973alw}
G.~'t Hooft,
Nucl. Phys. B \textbf{72}, 461 (1974)
doi:10.1016/0550-3213(74)90154-0

\bibitem{tHooft:1974pnl}
G.~'t Hooft,
Nucl. Phys. B \textbf{75}, 461-470 (1974)
doi:10.1016/0550-3213(74)90088-1

\bibitem{Frasca:2019ysi}
M.~Frasca,
Eur. Phys. J. C \textbf{80}, no.8, 707 (2020)
doi:10.1140/epjc/s10052-020-8261-7
[arXiv:1901.08124 [hep-ph]].

\bibitem{Frasca:2021yuu}
M.~Frasca, A.~Ghoshal and S.~Groote,
[arXiv:2109.05041 [hep-ph]].

\bibitem{Frasca:2013tma}
M.~Frasca,
Eur. Phys. J. C \textbf{74}, 2929 (2014)
doi:10.1140/epjc/s10052-014-2929-9
[arXiv:1306.6530 [hep-ph]].




\end{thebibliography}
\end{document}